\documentclass[aps,pra,superscriptaddress,twocolumn,twoside,a4paper,showpacs,floatfix]{revtex4}

\pdfoutput=1

\usepackage{hyperref,graphicx,amsmath,latexsym,revsymb,amssymb,verbatim,color}

\newcommand{\tr}{\textrm{tr}}
\newcommand{\bra}[1]{\langle #1|}
\newcommand{\ket}[1]{|#1\rangle}

\newcommand{\qh}{\frac{1}{\sqrt{2}}}

\newenvironment{proof}[1][Proof]{\begin{trivlist}
\item[\hskip \labelsep {\bfseries #1}]}{\end{trivlist}}

\begin{document}

\title{Demonstrating genuine multipartite entanglement and nonseparability without shared reference frames}
%Violating multipartite Bell inequalities without reference frames}

\author{Celal Furkan Senel}
\author{Thomas Lawson}
\author{Marc Kaplan}
\author{Damian Markham}
\author{Eleni Diamanti}
\affiliation{LTCI, CNRS -- T\'el\'ecom ParisTech, Paris, France}

\date{\today}
\begin{abstract}
Multipartite nonlocality is of great fundamental interest and constitutes a useful resource for many quantum information protocols.
However, demonstrating it in practice, by violating a Bell inequality, can be difficult.
In particular, standard experimental setups require the alignment of distant parties' reference frames, which can be challenging or impossible in practice. In this work we study the violation of certain Bell inequalities, namely the Mermin, Mermin-Klyshko and Svetlichny inequalities, without shared reference frames, when parties share a Greenberger-Horne-Zeilinger (GHZ) state. Furthermore, we analyse how these violations demonstrate genuine multipartite features of entanglement and nonlocality. For 3, 4 and 5 parties, we show that it is possible to violate these inequalities with high probability, when the parties choose their measurements from the three Pauli operators, defined only with respect to their local frames.
Moreover, the probability of violation, and the amount of violation, are increased when each party chooses their measurements from the four operators describing the vertices of a tetrahedron. We also consider how many randomly chosen measurement directions are needed to violate the Bell inequalities with high probability.
We see that the obtained levels of violation are sufficient to also demonstrate  \emph{genuine multipartite entanglement} and \emph{nonseparability}.
Finally, we show analytically that choosing from two measurement settings per party is sufficient to demonstrate the maximum degree of genuine multipartite entanglement and nonseparability with certainty when the parties' reference frames are aligned in one direction so that they differ only in rotations around one axis.
\end{abstract}

\maketitle

%%%%%%%%%%%%%%%%%%%%%%%%%%%%%%%%%%%%%%%%%%%%%%%%%%%%%%%%%%%%%%%%%%

Nonlocality is one of the most intriguing features of quantum mechanics. Testing it in the laboratory has therefore been the subject of great research efforts in the last few years. The motivations are twofold: witnessing nonlocality answers deep questions in fundamental physics; and it has practical applications such as generating random numbers \cite{Colbeck2011,Colbeck2012} and demonstrating secure communication \cite{Acin2006}.
But testing it, by violating a Bell inequality \cite{Bell1964}, is difficult in practice.
Two related problems are aligning the reference frames and calibrating the measurement devices of each party,
%For instance, in the lab parties using nonlocal states may struggle to align their reference frames or properly calibrate their measurement devices,
which can prevent the observation of nonlocality or, worse, lead to an erroneous certification of its existence \cite{Gisin2012}.

Recently, several works have addressed this issue in various settings.
For two parties, Shadbolt \emph{et al.} \cite{Shadbolt2011}, and independently Wallman \emph{et al.} \cite{Bartlett2012}, showed that nonlocality can be proven between two parties with no shared reference frame if each party measures the three Pauli operators; given a Bell state, a subset of the results of these measurements will almost certainly be able to violate the Clauser-Horne-Shimony-Holt (CHSH) inequality \cite{Clauser1969}.
Furthermore, Shadbolt \emph{et al.} \cite{Shadbolt2011} found the probability of a Bell violation when each party performs three or more random measurements, while Refs.~\cite{Bartlett2010, Bartlett2011} did the same for two randomly chosen measurements per party.
For more than two parties, Refs.~\cite{Bartlett2010, Bartlett2011,Bartlett2012} extended this idea to a class of multipartite Bell inequalities. In Ref.~\cite{Bartlett2012} it was shown numerically that the $n-$partite Mermin-Klyshko inequalities \cite{Mermin1990, Klyshko1993} (there referred to as the Mermin-Ardehali-Belinskii-Klyshko inequalities) are violated with certainty (or almost certainty for $n=3$) when each party measures the Pauli operators. 

In certain situations, some, but not total, alignment is possible between parties.
Indeed, the assumption that each party shares a single common axis is a standard noise model, used for instance in polarization, time and path encoded photonics \cite{Laing2010}. Ref.~\cite{Bartlett2011} showed analytically that $n$ parties whose reference frames are aligned in one direction can always violate the Mermin-Klysko inequalities by measuring a pair of operators in the plane orthogonal to the shared axis.

In the multipartite setting one is often interested not just in the existence of nonlocality and entanglement, but also one would like to confirm that it has features which are genuinely multipartite. That is, if $n$ parties share a state, which is entangled or nonlocal, they would like to know if this comes simply from two party properties - for example two out of the $n$ sharing an EPR state - or really from a property that is shared across all systems - known as genuine nonlocality and entanglement.

In this work, we investigate genuine multipartite features of entanglement and nonlocality in the absence of shared reference frames.
To do this, we extend the results of Refs.~\cite{Shadbolt2011, Bartlett2012, Bartlett2011} making use of three classes of multipartite Bell inequalities, namely the Mermin \cite{Mermin1990}, Mermin-Klyshko \cite{Mermin1990, Klyshko1993} and Svetlichny \cite{Svetlichny1986} inequalities, for parties sharing a Greenberger-Horne-Zeilinger (GHZ) state.
First, we numerically calculate not only the probability of violating each inequality when each party measures their arbitrarily rotated qubit with the Pauli operators, as in Refs.~\cite{Bartlett2010, Bartlett2011,Bartlett2012}, but also its value. Next we show that four measurement operators arranged as the vertices of a tetrahedron improve the probability and value of violation. We also consider how many randomly chosen measurement operators are needed to give a violation with high probability.
We show that the the level of violation is sufficient to demonstrate \emph{genuine multipartite entanglement} and \emph{separability} with high, almost certain, probability.
Finally, we consider the case where parties share one common axis only, and show that nonlocality, but also \emph{genuine multipartite entanglement} and \emph{separability}, can be demonstrated with certainty in this case.
We note that our analytical results for the Mermin inequalities replicate some of the results shown in Ref.~\cite{Bartlett2011}, using a different method. We go beyond this work by providing new analytical bounds for the Svetlichny inequalities in the odd $n$ case and the Mermin inequalities in the even $n$ case; furthermore, our analysis focuses on the genuine multipartite features demonstrated rather than the simple existence of nonlocality.

%%%%%%%%%%%%%%%%%%%%%%%%%%%%%%%%%%%%%%%%%%%%%%%%%%%%%%%%%%%%%%%%%%

\section{Bell inequality violation without reference frames in the bipartite case}
%\emph{Bell inequality violation without reference frames in the bipartite case.--}

Bell inequalities test whether the behavior of a quantum system is described by a local hidden variable (\emph{lhv}) theory, whereby a system acts according to a predetermined local deterministic strategy (or a probabilistic mixture of such strategies). In a \emph{lhv} model, the probability of measurements $A_1$ and $A_2$ on each half of a bipartite quantum state yielding results $\nu_1$ and $\nu_2$ is
\begin{align}\label{eq: lhv prob}
 P(\nu_1, \nu_2) = \int d \lambda \Delta(\lambda) P(\nu_1 | A_1, \lambda) P(\nu_2 | A_2, \lambda),
\end{align}
where $P(\nu_i | A_i, \lambda)$ is the probability of operator $A_i$ giving result $\nu_i$, and where $\lambda$ is a \emph{local hidden variable}, occurring with probability $\Delta(\lambda)$. It is known that \emph{lhv} models cannot account for the predictions of quantum mechanics~\cite{Bell1964}. The most famous illustration is provided by the CHSH inequality; according to \emph{lhv} models, the expectation values of measurements on a bipartite quantum state respect $\mathcal{CHSH} \leq1$, where
\begin{align}
\mathcal{CHSH} := \frac{1}{2} | E(A_1 A_2) + E(A'_1 A_2) + E(A_1 A'_2) - E(A'_1 A'_2) |,
\end{align}
and the expectation values, $E(A_1 A_2)$, are calculated based on Eq. \eqref{eq: lhv prob}. This inequality can be violated, \emph{i.e.} $\mathcal{CHSH} > 1$, using quantum mechanical expectation values, such as $E(A_1 A_2)= \tr(A_1 \otimes A_2 \ket{\phi^-} \bra{\phi^-})$, where the maximally entangled state $\ket{\phi^-} = (\ket{01} - \ket{10})/\sqrt{2}$ is measured using single qubit observables $A_1$ and $A_2$.

Shadbolt \emph{et al.} \cite{Shadbolt2011}, Liang \emph{et al.} \cite{Bartlett2010} and Wallman \emph{et al.} \cite{Bartlett2012} considered violating the CHSH inequality between two parties who do not share a \emph{global} reference frame. In this case, the quantum state can be written as $\overline{\rho} = (R_1 \otimes R_2)  \ket{\phi^-} \bra{\phi^-} (R_1 \otimes R_2)^{\dagger}$, where $\overline{\rho}$ denotes that the state $\ket{\phi^-}$  has undergone arbitrary unknown local rotations
\begin{align}\label{eq: Rj}
R_j = \cos \frac{\theta_j}{2}  \mathbb{I} - i \sin \frac{\theta_j}{2}  ( n_j^1 \sigma_1 + n_j^2 \sigma_2 + n_j^3 \sigma_3),
\end{align}
where $\theta_j$ and $n_j^k$ are real, $\sum_k n_j^k =1$, and $\sigma_1= \ket{1}\bra{0}+ \ket{0}\bra{1}$, $\sigma_2=  i \ket{0}\bra{1} -i \ket{1}\bra{0}$, $\sigma_3= \ket{0}\bra{0} - \ket{1}\bra{1}$ are the Pauli operators.
Since the rotations are unknown, this state can alternatively be thought of as a mixed state integrated over all values of $R_j$ \cite{Bartlett2007}. However, the form of Eq. \eqref{eq: Rj} suffices for our analysis since our results will be independent of $R_j$.

%%%%%%%%%%%%%%%%%%%%%%%%%%%%%%%%%%%%%%%%%%%%%%%%%%%%%%%%%%%%%%%%%%

\section{Multipartite Bell inequalities}
%\emph{Multipartite Bell inequalities.--}
We consider three classes of $n$-party Bell inequalities:
the Mermin ($\mathcal{M}$) \cite{Mermin1990},
the Mermin-Klyshko ($\mathcal{MK}$) \cite{Mermin1990, Klyshko1993},
and the Svetlichny ($\mathcal{S}$) inequalities \cite{Svetlichny1986}.
As in the CHSH inequality, each party $k$ performs measurements in two bases, $A_k$ or $A_k'$, which give outcomes $a_k\pm1$. According to a \emph{lhv} model the bound of each inequality is $1$, however some entangled states can violate this bound. We consider the  $n-$party GHZ state, $\ket{G_n} = ( \ket{0}^{\otimes ^n}+ \ket{1}^{\otimes ^n})/\sqrt{2} $, which maximally violates these Bell inequalities.

We use a standard formulation for constructing the Bell inequalities \cite{Collins2002}, according to which a Bell \emph{inequality} is made from a Bell \emph{polynomial}, $B$, which contains products such as $a_1^{(')} a_2^{(')}$. This is transformed into a Bell \emph{expression}, $\mathcal{B}$, by replacing these products with expectation values, $E(A_1^{(')} A_2^{(')})$, and taking the absolute value of the resulting expression. This is called the Bell \emph{value}. Finally, the Bell \emph{inequality} is constructed by introducing a bound respected by all \emph{lhv} models to the Bell expression, $\mathcal{B}\leq 1$. (Note that we will sometimes call $\mathcal{B}$ the Bell inequality. The bound $\mathcal{B}\leq 1$ is implied.)
As an example, in the bipartite case the CHSH polynomial is
\begin{align}
CHSH := \frac{1}{2} \big( a_1 a_2 + a'_1 a_2 + a_1 a'_2 - a'_1 a'_2 \big).
\end{align}

In the multipartite case, nonlocality alone is not the only quality that one can test with a Bell inequality.
There are two other interesting properties.
First, we can detect \emph{genuine multipartite entanglement}, GME($m$), where a state of $m$ systems is said to have genuine $m-$party entanglement if there exists no bipartite cut across which it is separable - that is all systems are involved in the entanglement. An $n$-party state is said to contain GME($m$) if some choice of $m$ subsystems has GME($m$).
%there is no separation into $n$ or fewer parties where it is not entangled.
Second, we can have a similar notion for nonlocality called \emph{separability}. We say that $n$ systems demonstrate Sep($l$) if there is no partitioning of the $n$ into more than $l$ groups such that they are local \cite{Collins2002} with respect to that partitioning. For example Sep($1$) means that if the $n$ systems are separated into two groups, they are nonlocal with respect to this partition (i.e. they would violate some two-party Bell inequality), for any such partitioning.

In general, nonlocality and entanglement are not identical, nonlocality being the stronger property. This means that there are states which are entangled but do not exhibit nonlocality, but the inverse is not true. This holds for the aforementioned multipartite notions as well; there exist states of $n$ systems which contain GME($n$) but do not demonstrate Sep($1$) nonlocality. That is, these states are entangled across any bipartition but for some bipartitions they do not violate any Bell inequality.

In addition to demonstrating the presence of some nonlocality, as studied in Refs.~\cite{Bartlett2012, Bartlett2010, Bartlett2011}, the Bell inequalities used in this work can detect these multipartite notions. The Mermin-Klyshko inequalities can differentiate different classes of genuine multipartite entanglement \cite{YLiang2014, Liang2014, Nagata2002, Yu2003}; as can the Svetlichny inequalities for separability \cite{Collins2002, Bancal2009}; finally, the Mermin inequalities have maximum algebraic values saturated by entangled quantum states.
%We consider the Mermin-Klyshko inequalities, which can prove GME($n$);
%the Svetlichny inequalities, which differentiate Sep($m$);
%and finally the Mermin inequalities, which have maximum algebraic values saturated by entangled quantum states.

%\subsection{Mermin-Klyshko Inequalities}
The Mermin-Klyshko expressions \cite{Mermin1990, Klyshko1993} are generated by the polynomials
\begin{align}
 MK_n := \frac{1}{2} MK_{n-1} (a_n + a_n') + \frac{1}{2} MK'_{n-1} (a_n - a_n'),
\end{align}
where $MK'_{k}$ is found by exchanging all $a_i$ and $a_i'$ in $MK_{k}$ \cite{Collins2002}.
The fundamental MK polynomial is the CHSH polynomial, $MK_2 = CHSH$.
The $n$-party $\mathcal{MK}$ inequality, $\mathcal{MK}_n \leq 1$, can be violated using entangled states.
For $n \leq 5$ -- as shall be the case for our investigation --,
if the state's largest entangled subspace contains no more than $m$ parties, where $1 \leq m \leq n$,
then $\mathcal{MK}_n\leq 2^{(m-1)/2}$ \cite{YLiang2014}; a violation of this bound implies GME($m+1$). %\cite{Collins2002}.
For instance, for three parties
\begin{align}
 MK_3 := \frac{1}{2} \big(a_1 a_2 a_3' + a_1 a_2' a_3  + a_1' a_2 a_3 - a_1' a_2' a_3' \big),
\end{align}
$\mathcal{MK}_3 > \sqrt{2}$ shows genuine three party entanglement.
Note that for arbitrary $n$ we can still characterize entanglement, albeit less precisely: all biseparable correlations satisfy $\mathcal{MK}_n\leq 2^{\frac{n}{2} -1}$  \cite{Liang2014}; conversely $\mathcal{MK}_n > 2^{\frac{n}{2} -1}$ implies the correlations are not biseparable, which in turn implies the state is GME($n$). Taking into account the number of unentangled \emph{particles} -- as well as entangled ones -- lets one further differentiate the entanglement classes \cite{Nagata2002, Yu2003}.

%\subsection{Svetlichny Inequalities}
The Svetlichny polynomials \cite{Svetlichny1986, Collins2002} are
\begin{align}
S_n := \frac{1}{2}(MK_n + MK'_n)
\end{align}
for $n$ odd, while they coincide with $MK_n$ for $n$ even.
The Svetlichny inequality is $\mathcal{S}_n \leq 1$.
For GME($n$) (corresponding to Sep(1)) $S_n \leq  2^{\frac{n-1}{2}}$ for $n$ even and $S_n \leq  2^{\frac{n-2}{2}}$ for $n$ odd.
If the state belongs to Sep($m$)
then $S_n \leq 2^{\frac{n-m}{2}}$ for $n$ even and $S_n \leq 2^{\frac{n-m-1}{2}}$ for $n$ odd \cite{Collins2002, Bancal2009}.
Considering a system of three parties we see that separability is \emph{a priori} different to entanglement.
A pair of generally nonlocal states locally connected to a third one (Sep(2)) gives $\mathcal{S}_3 \leq 1$; a three party entangled state (GME(3)) can violate this, reaching $\mathcal{S}_3 = \sqrt{2}$ \cite{Svetlichny1986}.

%\subsection{Mermin Inequalities}
Finally, the Mermin inequalities are derived from the polynomials \cite{Mermin1990}
\begin{align}
 M_n := & \frac{1}{2^{\frac{n+2}{2}}i} \prod_{j=1}^n \big(a_j + i a_j' \big) \notag\\
 - & \frac{1}{2^{\frac{n+2}{2}}i}  \prod_{j=1}^n \big(a_j - i a_j' \big),
\end{align}
for $n$ even. ($M_n$ coincides with $MK_n$ for $n$ odd.)
In other words $M_n$ contains all permutations of $l$ primed and $n-l$ unprimed operators, where $l$ is odd.
Terms $l=3, 7, 11, \ldots$ have a coefficient $-1$.
The Mermin inequality is $\mathcal{M}_n \leq 1$.
For the state $\ket{G_n}$, $\mathcal{M}_n \leq 2^{\frac{n}{2}-1}$,
% coefficients $N_{odd} = \frac{1}{2^{\frac{n+1}{2}}}$ and $N_{even} = \frac{1}{2^{\frac{n+2}{2}}}$ We consider just $n$ eve so ok. FINE!
the algebraic maximum of $\mathcal{M}_n$, for odd $n$. This means that no combination of expectation values taking the values $\pm1$ can outperform the predictions of quantum mechanics.

%**************************************************
\begin{figure}[tb]
\centering
\includegraphics[scale=0.07]{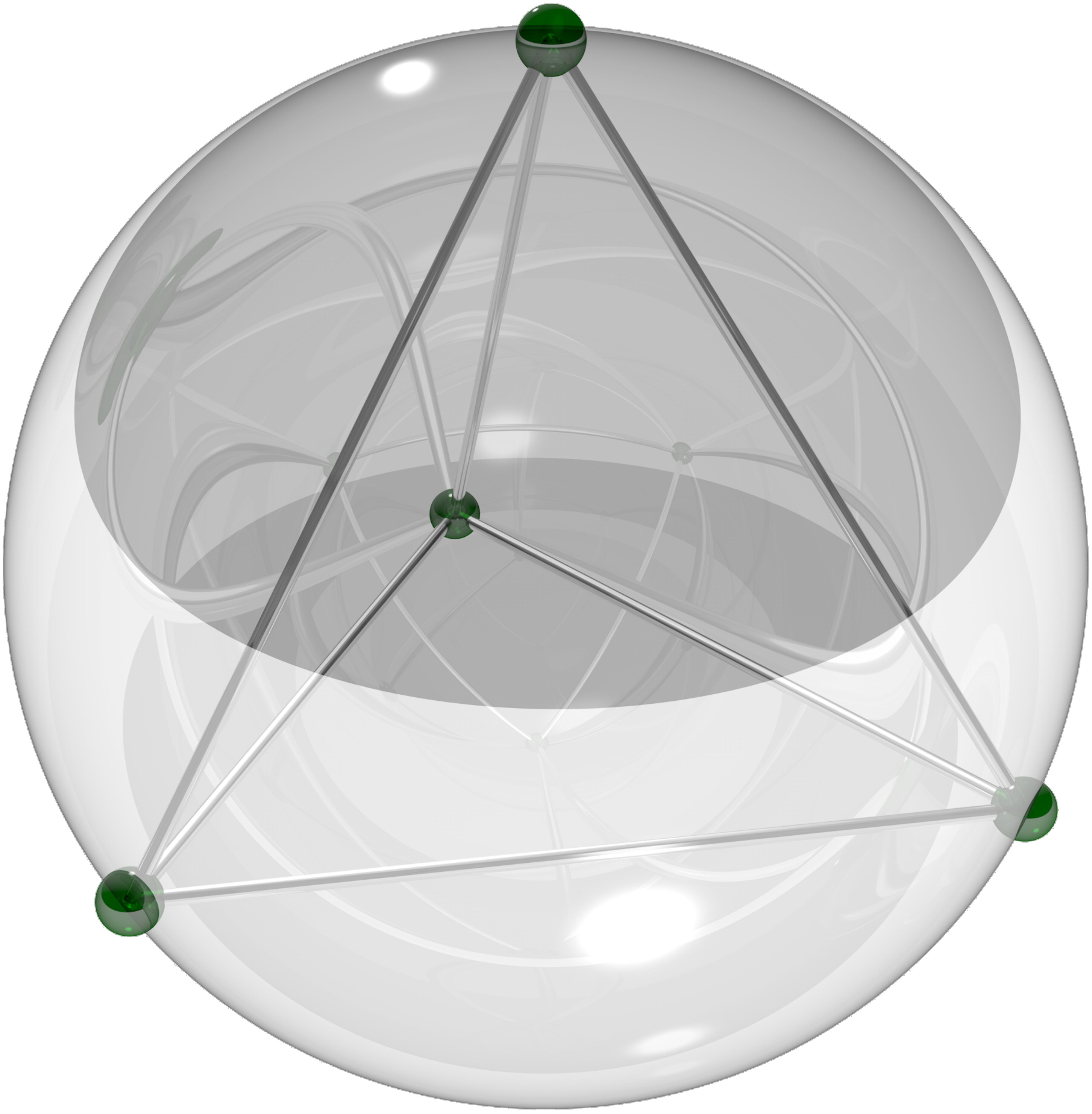}
\caption{The four vertices of the tetrahedron are used to define measurement directions evenly spaced over the Bloch sphere.}
\label{fig: Tet}
\end{figure}
%**************************************************

\section{Results for general rotations}
%\emph{Results for general rotations.--}
We now consider the case where $n$ parties share a GHZ state, but do not share a \emph{global} reference frame. Equivalently, each part of $\ket{G_n}$ undergoes a random local rotation, $\ket{\overline{G}_n} = (R_1 \otimes R_2 \otimes \ldots \otimes R_n )\ket{G_n}$, with $R_n$ chosen according to the Haar measure.

%**************************************************
\begin{figure*}[tb]
\centering
\includegraphics[scale=1.4]{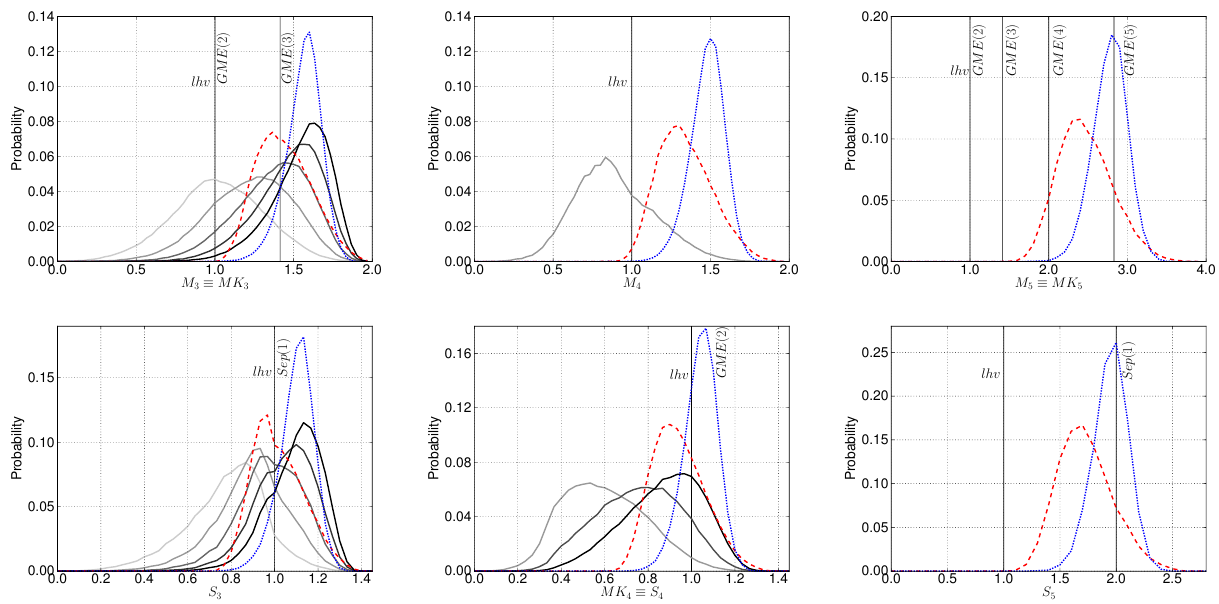}
\caption{Probability distributions of $\mathcal{M}_n$, $\mathcal{MK}_n$, and $\mathcal{S}_n$ values for $n = 3, 4, 5$ parties when a randomly rotated GHZ state is measured in the Pauli (red, dashed) and tetrahedral (blue, dotted) bases (color online), and in $3, 4, 5, 6, 7$ randomly chosen directions (in increasingly dark shades of grey). The vertical lines mark the \emph{lhv}, and multipartite entanglement/separability bounds. Note that the latter calculation was not possible for all numbers of random measurement operators beyond the three party case.}
\label{fig: results}
\end{figure*}
%**************************************************

We numerically calculate the probability of violating the $\mathcal{M}_n$, $\mathcal{MK}_n$, and $\mathcal{S}_n$ inequalities for $n = 3, 4, 5$ parties for the following protocol:
each party measures his share of $\ket{\overline{G}_n}$ in a number of bases.
We then pick the two bases from each party whose results maximize each Bell value.
We consider three settings.
First, we assume the parties have \emph{local} reference frames.
Each party measures the Pauli operators, $\pm \sigma_i$, on $\ket{\overline{G}_n}$. As we will show, three measurement operators are not always enough to guarantee a Bell violation.
The second setting uses four measurement directions -- defined by the vertices of a tetrahedron, Fig. \ref{fig: Tet} -- per qubit to improve the probability of violation.
In the third setting, we assume that the parties have neither a global nor a local reference frame and calculate the distribution of Bell values when each party measures in a number of random directions, chosen according to the Haar measure.
We want to know how many measurement operators we need to observe a violation with high probability.

Fig. \ref{fig: results} shows our results.
We make several observations.
First we consider the Pauli operators which often, but not always, suffice to violate the $\mathcal{M}_n$ and $\mathcal{MK}_n$ inequalities: the $\mathcal{M}_3$ inequality is violated with probability roughly $99.99\%$. %, according to reference \cite{Bartlett2012}.
An example of a rotated state that does not allow a violation is $\overline{\ket{G_3}} = R_t \otimes R_t \otimes R_t \ket{G_3}$, where
\begin{align}
R_t = \cos \frac{\theta_t}{2}  \mathbb{I} - i \sin \frac{\theta_t}{2}  \qh (\sigma_1 + \sigma_2),
\end{align}
and $\theta_t = \arctan \sqrt{2}$. This rotation gives the three observables an equal component on the $\sigma_1-\sigma_2$ plane. In this case $\mathcal{M}_3 = 0.98$, implying that there exists a set of states of non-zero measure containing $\overline{\ket{G_3}}$ such that $\mathcal{M}_3<1$.
For $n = 5$ parties, the Pauli operators %appear to
show GME(3) with certainty, while the GME(5) and Sep(1) bounds can also be violated, albeit with lower probabilities (around $19\%$ for GME(5) and $18\%$ for Sep(1)).

The tetrahedral basis gives better results for all inequalities, violating the Mermin inequalities with almost unit probability, although there is still a chance of not violating -- around $10^{-5}$ for the $\mathcal{M}_3$ inequality. One non-violating rotated state is $\overline{\ket{G_3}} = \mathbb{I}\otimes \mathbb{I} \otimes R_s \ket{G_3}$, where $R_s = \cos (3 \pi/20) \mathbb{I} - i \sin (3 \pi/20) \sigma_1$, giving $\mathcal{M}_3 = 0.93$.
The tetrahedral basis also improves the probability of demonstrating genuine multipartite entanglement: for $n = 3$ parties, GME(3) is demonstrated with a probability close to $92\%$.
As in the bipartite case \cite{Shadbolt2011}, random measurements are less effective; even with seven operators, the $\mathcal{M}_3$ inequality is only violated with probability $81\%$.

For four parties we see that the tetrahedral basis gives violation of $\mathcal{M}_4$ with almost certainty, hence demonstrating non-locality, although we do not see genuine multipartite features here. For five parties we see that the tetrahedral basis gives almost certain demonstration of non-locality, and further that the state demonstrates Sep(2) and that it contains GME(4) with high probability.

Finally, we note that the Svetlichny inequalities are harder to violate than the Mermin and Mermin-Klyshko inequalities; for instance, the $\mathcal{S}_3$ inequality is violated with probability roughly $55\%$ using the Pauli operators. This is because these inequalities test nonseparability, which is more general than entanglement.

%%%%%%%%%%%%%%%%%%%%%%%%%%%%%%%%%%%%%%%%%%%%%%%%%%%%%%%%%%%%%%%%%%%%%%%%

\section{Restricted rotations}
%\emph{Results for restricted rotations.--}
We now consider the special case of rotations restricted to the $\sigma_1-\sigma_2$ plane,
 $\ket{\overline{G}_n} = (R^z_1 \otimes R^z_2 \otimes \ldots \otimes R^z_n ) \ket{G_n}$, where
\begin{align}
 R^z_i = \cos \frac{\theta_i}{2} \mathbb{I} - i \sin\frac{\theta_i}{2} \sigma_3.
 \end{align}
We show analytically that two perpendicular measurement operators on that plane are sufficient to violate the $\mathcal{M}_n$ and the $\mathcal{S}_n$ inequalities with certainty.
Furthermore, the violation is always large enough to demonstrate full genuine multipartite entanglement, GME($n$), for $\mathcal{M}_{n_{odd}}$ and complete nonseparability, Sep(1), for $\mathcal{S}_{n}$.
Note that the identities $\mathcal{M}_{n_{odd}} \equiv \mathcal{MK}_{n_{odd}}$ and  $\mathcal{S}_{n_{even}} \equiv \mathcal{MK}_{n_{even}}$
mean this section has some overlap with Ref.~\cite{Bartlett2011}, which lowerbounded the value of $\mathcal{MK}_{n}$ in this scenario, although the authors did not consider genuine multipartite entanglement and separability.
Indeed, we use the result of Ref.~\cite{Bartlett2011} to show that $\mathcal{MK}_{n_{even}}$ always demonstrates GME($n$) and, when interpreted as $\mathcal{S}_{n_{even}}$, also demonstrates Sep(1).
For $\mathcal{M}_{n_{odd}}$, we derive using our proof that GME($n$) is demonstrated with certainty; however, the same observation could have been made given the result of Ref.~\cite{Bartlett2011}.

First, let us consider the $n-$party Mermin inequalities.
\begin{proof}
Choosing operators $A_i =\sigma_1$ and $A'_i =\sigma_2$ the expectation values are
\begin{align}
E(A^{(')}_1 A^{(')}_2 \ldots A^{(')}_n) = \cos \big( \Theta - p \frac{\pi}{2}\big),
\end{align}
where $\Theta= \sum_i \theta_i$ and $p$ is the number of primed terms.
The Bell value is
\begin{align}\label{eq: bell value cos}
\mathcal{M}_n =  N |  \sum_{p=1, 5, 9 \ldots} & {n \choose p} \cos \big( \Theta - p \frac{\pi}{2}\big)\notag\\
- \sum_{p=3, 7, 11 \ldots} & {n \choose p} \cos \big( \Theta - p \frac{\pi}{2}\big) |,
\end{align}
where $N= 2^{-\frac{n}{2}}$ for $n$ even and $N= 2^{-\frac{n-1}{2}}$ for $n$ odd.
The terms $p=1, 5, 9, \ldots$ are $\sin\Theta$. Terms $p=3,7,11, \ldots$ have a minus sign, $-\sin\Theta$.
The Bell value, $\mathcal{M}_n =  N \sum_{p \text{ odd}} {n \choose p} |\sin \Theta|
= N 2^{n-1} |\sin \Theta|$, depends on whether $n$ is odd or even. First, we consider $n$ odd,
\begin{align}
\mathcal{M}_{n_{odd}}  = 2^{\frac{n -1}{2}} |\sin \Theta|.
\end{align}
The inequality is violated whenever
\begin{align}\label{eq: rot z cond viol}
| \sin \Theta | > \frac{1}{2^{\frac{n -1}{2}}}.
\end{align}
This is not satisfied for some values of $\Theta$ but, in this case, a different combination of measurements will give a violation.
Swapping the operators, so that $A_i =\sigma_2$ and $A'_i =\sigma_1$, gives expectation values
\begin{align}
E(A^{(')}_1 A^{(')}_2 \ldots A^{(')}_n)
&= \cos \big( \Theta - (n-p) \frac{\pi}{2}\big). \notag\\
\end{align}
For $n$ odd $n-p$ is even, so the expectation values are $\cos \Theta $ for $n-p = 0, 4, 8, \ldots$ and $-\cos \Theta $ for $n-p = 2, 6, 10, \ldots$,
giving $\mathcal{M}_{n_{odd}} = 2^{\frac{n -1}{2}} |\cos \Theta|$.
Hence, for $n$ odd, the Bell value is $\mathcal{M}_{n_{odd}} \geq \max \lbrace{ 2^{\frac{n -1}{2}} | \sin \Theta |, 2^{\frac{n -1}{2}} | \cos \Theta | \rbrace} \geq 2^{\frac{n}{2} -1}$. %, which is always greater than $\sqrt{2}$, since the smallest odd $n$ is 3.

For $n$ even, using $A_i =\sigma_1$ and $A'_i =\sigma_2$, we have
\begin{align}
\mathcal{M}_{n_{even}}  = 2^{\frac{n}{2}-1} |\sin \Theta|.
\end{align}
For values of $\Theta$ for which the inequality is not violated we set $A_1=\sigma_2$, $A'_1 = - \sigma_1$, $A_{i>1}=\sigma_1$ and $A'_{i>1} = \sigma_2$
($- \sigma_1$ is done by flipping the sign of any expectation value containing $\sigma_1$)
giving $\mathcal{M}_{n_{even}}  = 2^{\frac{n}{2} -1} |\cos \Theta|$.

Since the Mermin and Mermin-Klyshko inequalities are equivalent for odd $n$, we can apply the genuine multipartite entanglement bounds to $M_{n_{odd}}$.
We see from our lowerbound, $M_{n_{odd}} > 2^{\frac{n}{2}-1}$, that $M_{n_{odd}}$ demonstrates GME($n$) with certainty for %$n \leq 5$ and biseparability of the quantum state, GME($\geq 2$),
arbitrary $n$.
\end{proof}

The proof for the Svetlichny inequalities follows a similar argument.
\begin{proof}
We consider odd $n$. Using the identity $M_{n_{odd}} \equiv MK_{n_{odd}}$, the Svetlichny polynomial becomes
\begin{align}
S_{n_{odd}} = \frac{1}{2} \big( M_{n_{odd}} + M_{n_{odd}}'\big).
\end{align}
As before we will need two measurement strategies. The first is $A_i =\sigma_1$ and $A'_i =\sigma_2$ giving
%expectation values
%\begin{align}
%E(A^{(')}_1 A^{(')}_2 \ldots A^{(')}_n) = \cos \big( \Theta - p \frac{\pi}{2}\big).
%\end{align}
%The
%Bell value is
\begin{align}
\mathcal{S}_{n_{odd}} \!\! = \!\!
\frac{N}{2} | \!\!\!\! \sum_{p = 1, 5, 9 \ldots} \!\!\!\! & {n \choose p} \Big( \cos \big( \Theta - p \frac{\pi}{2}\big) + \cos \big( \Theta - (n-p)\frac{\pi}{2}\big) \Big)\notag\\
- \!\!\!\! \sum_{p=3, 7, 11 \ldots} \!\!\!\! & {n \choose p} \Big( \cos \big( \Theta - p \frac{\pi}{2}\big) + \cos \big( \Theta - (n-p)\frac{\pi}{2}\big) \Big)|,
\end{align}
where $N= 2^{-\frac{n-1}{2}}$.

We consider two cases. First, $n = 1, 5, 9 \ldots$, in which case
\begin{align}
\mathcal{S}_{n=1, 5, 9 \ldots} = \frac{N}{2} \sum_{p \text{ odd}}  {n \choose p} | \sin \Theta + \cos \Theta |.
\end{align}

Second, $n = 3, 7, 11 \ldots$, giving
\begin{align}
\mathcal{S}_{n=3, 7, 11 \ldots} = \frac{N}{2} \sum_{p \text{ odd}}  {n \choose p} |\sin \Theta - \cos \Theta |.
\end{align}

In both cases there are values of $\Theta$ for which the inequality is not violated. The second measurement strategy -- $A_1=\sigma_2$, $A'_1 = - \sigma_1$, $A_{i>1}=\sigma_1$ and $A'_{i>1} = \sigma_2$ -- introduces an additional rotation of $\pi/2$, hence,
\begin{align}\label{eq: Sn restricted}
\mathcal{S}_{n_{odd}} = & 2^{\frac{n-3}{2}}  \max \lbrace  | \sin \Theta \pm \cos \Theta |, \notag\\
& | \sin (\Theta + \frac{\pi}{2}) \pm \cos (\Theta+ \frac{\pi}{2}) | \rbrace.
\end{align}

Equation \eqref{eq: Sn restricted} implies $ \mathcal{S}_{n_{odd}} \geq 2^{\frac{n-3}{2}}$, which means that $ \mathcal{S}_{n_{odd}}$ demonstrates Sep(1) with certainty.
\end{proof}

For even $n$ the Svetlichny inequality coincides with the Mermin-Klyshko inequality, and so is covered by the result of Ref.~\cite{Bartlett2011} which shows that $\mathcal{MK}_{n_{even}} \geq 2^{\frac{n}{2} -1}$. In other words $\mathcal{S}_{n_{even}} \geq 2^{\frac{n}{2} -1}$, demonstrating Sep(1) with certainty. Furthermore, this result also implies that $\mathcal{MK}_{n_{even}}$ demonstrates GME($n$) with certainty.

%This restricted rotation is a common noise model used when one degree of freedom is stable, in photonic polarization, time, or path encoding, for instance .\cite{Laing2010}

%%%%%%%%%%%%%%%%%%%%%%%%%%%%%%%%%%%%%%%%%%%%%%%%%%%%%%%%%%%%%%%%%%%%%%

\section{Discussion}
%\emph{Discussion.--}
We have numerically calculated the probabilities of violating the Mermin, Mermin-Klyshko and Svetlichny inequalities, for three, four and five parties, when each party measures the Pauli operators on their share of a locally rotated GHZ state, showing that it is possible to reliably demonstrate multipartite nonlocality without a global reference frame.
The probability of violation increases when a tetrahedral basis of four operators is used, although even in this case violation is not guaranteed.
Increasing the number of measurement operators may help in this direction. The set of platonic solids, to which the tetrahedron belongs, is a natural way of distributing the measurement directions.
Several random measurement directions can also be used to violate the inequalities, albeit with lower probability.

In contrast with previous investigations, we have considered the genuine multipartite entanglement and separability. Being stronger than the nonlocality demonstrated by violating a Bell inequality, these concepts need a large violation of the Bell inequalities which becomes easier as $n$ increases, as noted in Ref.~\cite{Bartlett2010}. We see that for three and five parties genuine multipartite features can be demonstrated with high probability.

Finally, when the rotations of the quantum state are restricted to a plane, we show analytically, for arbitrary $n$, that just two operators per party are sufficient to violate the $ \mathcal{M}_{n}$ and $\mathcal{S}_{n}$ inequalities, and to demonstrate the maximal degree of genuine multipartite entanglement, GME($n$), and nonseparability, Sep(1), with certainty.

It will be interesting to study the possibility to demonstrate strong nonlocal phenomena for multipartite states other than the GHZ state.

Our results have a practical application since local state rotations can be thought of as a type of noise. In particular, the restricted case, where each party shares a common axis, is a common noise model for phonic quantum systems encoded in polarization, time or path \cite{Laing2010}.
In this sense, we have shown that highly sophisticated multipartite nonlocal phenomena can be seen even in the presence of noise.

%%%%%%%%%%%%%%%%%%%%%%%%%%%%%%%%%%%%%%%%%%%%%%%%%%%%%%%%%%%%%%%%%%%%%%%%

\vspace{0.2cm}
\noindent
\emph{Acknowledgments.--} We thank Y.-C. Liang, S. D. Bartlett and J. J Wallman for helpful comments. We acknowledge financial support from the French National Research Agency through the project 12-PDOC-0022-01NLQCC and the City of Paris through the project CiQWii. T.L. acknowledges support from Digiteo.\\

%%%%%%%%%%%%%%%%%%%%%%%%%%%%%%%%%%%%%%%%%%%%%%%%%%%%%%%%%%%%%%%%%%%%%%%%f

\bibliography{nonlocality}

%%%%%%%%%%%%%%%%%%%%%%%%%%%%%%%%%%%%%%%%%%%%%%%%%%%%%%%%%%%%%%%%%%%%%%%%

\end{document}